\documentclass[pra,
a4paper,
showpacs,
twocolumn,
superscriptaddress]{revtex4-1}
\usepackage{amssymb}
\usepackage{amsmath}
\usepackage{amsfonts}
\usepackage{braket}
\usepackage{graphicx}
\usepackage{bm}
\usepackage{color}
\usepackage{multirow}
\usepackage{natbib}
\usepackage{hyperref}
\usepackage[normalem]{ulem}
\usepackage{verbatim}
\def \be {\begin{equation}}
\def \ee {\end{equation}}
\def \bea {\begin{align}}
\def \eea {\end{align}}
\def \p {\partial}
\def \BEA {\begin{eqnarray}}
\def \EEA {\end{eqnarray}}

\def \BC {\begin{cases}}
\def \EC {\end{cases}}

\def \be {\begin{equation}}
\def \ee {\end{equation}}
\def \bea {\begin{align}}
\def \eea {\end{align}}
\def \p {\partial}
\def \BEA {\begin{eqnarray}}
\def \EEA {\end{eqnarray}}

\def \BC {\begin{cases}}
\def \EC {\end{cases}}
\usepackage{graphicx}
\usepackage{dcolumn}
\usepackage{bm}
\usepackage{amssymb}
\usepackage{amsmath}
\usepackage{verbatim}
\usepackage{color}
\usepackage{ulem}
\def \be {\begin{equation}}
\def \ee {\end{equation}}
\def \bea {\begin{align}}
\def \eea {\end{align}}
\def \p {\partial}
\def \BEA {\begin{eqnarray}}
\def \EEA {\end{eqnarray}}

\def \BC {\begin{cases}}
\def \EC {\end{cases}}

\newcommand{\aleq}[1]{
\begin{equation}
    \begin{aligned}
    #1
    \end{aligned}
\end{equation}
}

\begin{document}
\title
{
 Spin and charge transport through  helical Aharonov-Bohm interferometer with strong magnetic impurity
}

\author{R.\,A.~Niyazov}
\address{Department of Physics, St. Petersburg State University, St. Petersburg 198504, Russia}
\address{NRC ``Kurchatov Institute'', Petersburg Nuclear Physics Institute, Gatchina 188300, Russia}
\author{D.\,N.~Aristov}
\address{NRC ``Kurchatov Institute'', Petersburg Nuclear Physics Institute, Gatchina 188300, Russia}
\address{Department of Physics, St. Petersburg State University, St. Petersburg 198504, Russia}
\affiliation{Ioffe Institute,
194021 St.~Petersburg, Russia}
\author{V.\,Yu.~Kachorovskii }
\affiliation{Ioffe Institute,
194021 St.~Petersburg, Russia}
\affiliation{CENTERA Laboratories, Institute of High Pressure Physics, Polish Academy of Sciences, 01-142 Warsaw, Poland
 }


\begin{abstract}
We  discuss 
transport through an interferometer formed by helical edge states of the quantum spin Hall insulator. Focusing on effects induced by a  strong magnetic impurity placed in one of the arms of interferometer, we  
consider the experimentally relevant case of relatively high temperature as compared to the level spacing. We obtain the conductance and the spin polarization in the closed form 
for arbitrary  tunneling amplitude of the contacts and arbitrary strength of the magnetic impurity. We demonstrate the existence of  quantum effects which do not show up in previously studied case of weak magnetic disorder. 
We find optimal conditions for spin filtering and demonstrate that the spin polarization of outgoing electrons can reach 100\%. 
\end{abstract}

\maketitle
\section{Introduction}

A novel class of materials --- topological insulators --- has become a hot topic in the last decade.  
These materials 
are  insulating in the bulk, but  exhibit conducting states at  the edges of the sample~\cite{Bernevig2013,Hasan2010,Qi2011}. The  edge states  demonstrate 
  surprising properties. In particular, in  two-dimensional (2D)  topological insulators (TI) edge states are one-dimensional channels where: (i) the electron spin
 is tightly connected to the electron motion direction, e.g. electrons with spin-up and spin-down propagate in opposite directions; (ii) the electron transport is ideal, in the sense that electrons do not experience backscattering from conventional non-magnetic impurities, similarly to what occurs in edge states of Quantum Hall Effect, but without invoking high magnetic fields.
  The 2D  topological insulator phase was predicted in HgTe quantum wells \cite{Kane2005,predicted} and confirmed by direct   measurements of conductance of the  edge states \cite{confirmed} and  by  the experimental analysis of the non-local transport   \cite{Roth2009,Gusev2011,Brune2012,Kononov2015}.
 
 Considerable attention was paid to the analysis of the Aharonov Bohm (AB) effect in 2D  TI: the dependence of the longitudinal conductance of  nanoribbons  and nanowires  on the  magnetic  flux  piercing   their cross-section    was studied   \cite{Peng2010,Lin2017}; weak antilocalization was  investigated in the disordered topological insulators  and oscillations with magnetic flux with the period   equal to the half of the flux quantum  were predicted \cite{Bardarson2010,Bardarson2013}.   The AB effect was  discussed for    almost closed loops formed by curved edge states   \cite{Buttiker2012}.  Also, the AB oscillations were  observed in the magnetotransport measurements of  transport (both local and nonlocal) in 2D topological insulators based on HgTe quantum wells \cite{KvonAB2015} and were explained  by coupling of helical edges to bulk puddles of charged carriers.

\begin{figure}[b]
\includegraphics[width=0.8\columnwidth]{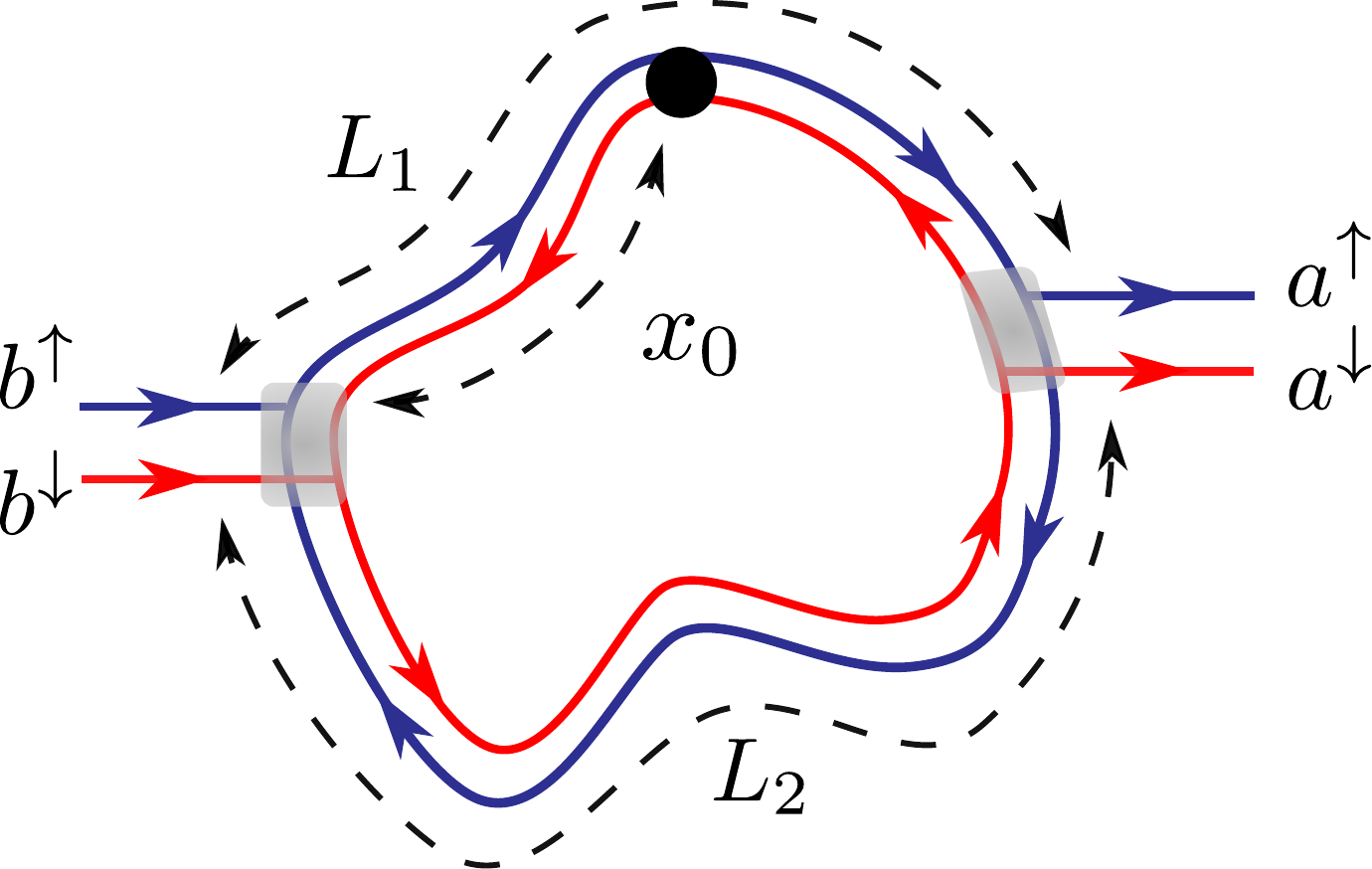}
\caption{\label{fig:transfermatrix}
Helical Aharonov-Bohm interferometer with the magnetic impurity placed in the upper shoulder.  }
\end{figure}

 The purpose of the current paper is to study standard AB setup  based on helical-edge states (HES)   of a quantum spin Hall insulator   tunnel-coupled
 to leads  (see Fig.~\ref{fig:transfermatrix}).  

 Such an  interferometer 
 was already studied theoretically at zero temperature  \cite{Chu2009, Masuda2012,Dutta2016}. Here, we  focus on 
 the case of relatively high temperature,
$ T\gg \Delta,$
where 
   $\Delta=2\pi v_F/L$ is the level spacing  which is controlled by total edge  length $L=L_1+L_2,$ where $L_{1,2}$ are lengths of the interferometer's shoulders. 
    For typical sample parameters, $L =10~ \mu$m  and $v_F=10^7$ cm/s, we estimate  the level spacing $\Delta \approx 3 $ K. 
As seen from this estimate, the case  $T \gg\Delta$ is interesting for possible applications. 
There is also upper limitation for  temperature. For good quantization,  $T$  should be much smaller than the bulk gap of the topological insulator: $T\ll \Delta_{\rm b}$.    For the first time quantum spin Hall effect was observed in structures based on HgTe/CdTe~\cite{Konig2007} and InAs/GaSb~\cite{Knez2011}, which had a rather narrow bulk gap, less than 100 K. Substantially large values were observed recently in WTe$_2,$ where gap of the order of 500 K was observed~\cite{Wu2018},  and in bismuthene grown on a SiC (0001) substrate, where a bulk gap of about 0.8 eV was demonstrated~\cite{Reis2017,Li2018} (see also recent discussion in Ref.~\cite{Stuhler2019}).    Thus, recent experimental studies unambiguously indicate the possibility of transport through HES at room temperature, when the condition 
\be 
\Delta_{\rm b} \gg T\gg \Delta  
\label{TggDelta}
\ee
can be easily satisfied.

    The high-temperature regime, $T\gg \Delta,$  was already studied for single-channel AB interferometers made of conventional  materials   \cite{jagla,dmitriev,Shmakov2013,Dmitriev2015,SCS2017} and it was demonstrated that  flux-sensitive interference   
    effects survive in this case.     Recently, we discussed  high-temperature  electron and spin transport in AB interferometers 
    based on helical edge states of TI \cite{Niyazov2018,Niyazov2020}.   We considered  setup shown in  Fig.~\ref{fig:transfermatrix}  and assumed  that there is a {\it weak}
  magnetic impurity (or {\it weak} magnetic disorder). We
 found that  both tunneling conductance $G$ and the spin polarization $\mathcal{P}$  of outgoing electrons  show  sharp resonances appearing  periodically with 
dimensionless  flux,
$\phi=\Phi/\Phi_0,$  with the period $\Delta \phi=1/2.$  
Here  
 $\Phi$ is the external    magnetic
flux piercing the area encompassed by  edge states and  $\Phi_0=hc/e$ is the flux quantum.
Simple estimates show that condition  $ \Phi \sim \Phi_0$, is achieved for interferometer with  HES of typical length  $L =10~ \mu$m   in fields  $B\sim 3\mbox{ Oe}$,  well below the expected magnitude of the fields destroying the  edge states
\cite{Du2015,Zhang2014,Hu2016a}. 

Importantly,   condition \eqref{TggDelta} ensures the universality of spin and charge transport (see 
discussion in Refs.~\cite{Niyazov2018,Niyazov2020}), which do not depend on details of the systems, in particular, on the device geometry.
  A very sharp dependence of the conductance and the spin polarization  on $\phi$,  predicted in  Refs.~ \cite{Niyazov2018,Niyazov2020},  is very promising  for applications  for  tunable spin filtering and in the area of  extremely sensitive detectors of magnetic fields.
 We also demonstrated that charge and spin transfer through the  AB  helical interferometer   can  be described in terms of the ensemble of the flux-tunable  qubits  \cite{Niyazov2020} that     opens a wide avenue  for high-temperature quantum computing.

 In this paper, we generalize  the  results obtained in Refs.~\cite{Niyazov2018,Niyazov2020}  for the case of a {\it strong } impurity.   We study electrical and spin  transport through  AB helical interferometer containing a single magnetic impurity of arbitrary strength and find optimal conditions for spin filtering.  
  We also demonstrate that with increasing of the  impurity  
 strength  new quantum processes come into play which do not show up  for {\it weak} impurity.  Most importantly, we confirm the idea which was put forward previously  \cite{Niyazov2020} but has not yet been verified by direct calculations. We demonstrate   that   a strong magnetic impurity inserted  into  one of the interferometer's  shoulder, blocks the transition through this shoulder and  only the other shoulder remains active.  As a result,  the spin polarization of outgoing electrons can achieve 100\%.  Remarkably, this mechanism is robust to dephasing by a non-magnetic bath, works at high temperatures and  thus 
 has high prospects in the quantum computing.

\section{Model}
We consider tunneling  charge and spin transport through  AB interferometer based on HES. 
 We limit ourselves to a discussion of  setup with  a single {\it strong} impurity placed into the upper shoulder at the distance $x_0$ (along the edge) from the left contact (see Fig.~\ref{fig:transfermatrix}).  
We discuss the dependence of the tunneling conductance $G$  and  spin polarization 
of  outgoing electrons $\mathcal{P}$ on the    
external  dimensionless  magnetic
flux $\phi$.  
Similar to Refs.~ \cite{Niyazov2018,Niyazov2020},  we neglect  the influence of the magnetic fields on the helical states.

We assume that 
the impurity is classical with large magnetic moment
 $\mathbf M$ ($M\gg 1$) and
 describe  such an impurity by the following
  scattering matrix
\be
\hat S_M= \begin{pmatrix}
  e^{i \zeta} \cos \theta &  i \sin \theta ~e^{i\varphi}\\
i \sin \theta ~e^{-i\varphi} & e^{-i \zeta} \cos \theta\\
\end{pmatrix}.
\label{SM}
\ee
One can show that the forward scattering phase $\zeta$ can be absorbed into the shift of $\phi$ and is put to zero  below. 
We neglect  feedback  effect related to the dynamics of $\mathbf M$ caused by  exchange  interaction with the ensemble of  right- and left-moving  electrons~\cite{kur3} assuming that the direction of  $\mathbf M$ is controlled, e.g.,  
by in-plane magnetic field, which does not affect the AB interference, or by the magnetic anisotropy of the impurity Hamiltonian. The scattering of electrons may also happen off the ferromagnetic tip placed in the vicinity of HES.

\begin{figure}
\includegraphics[width=0.5\columnwidth]{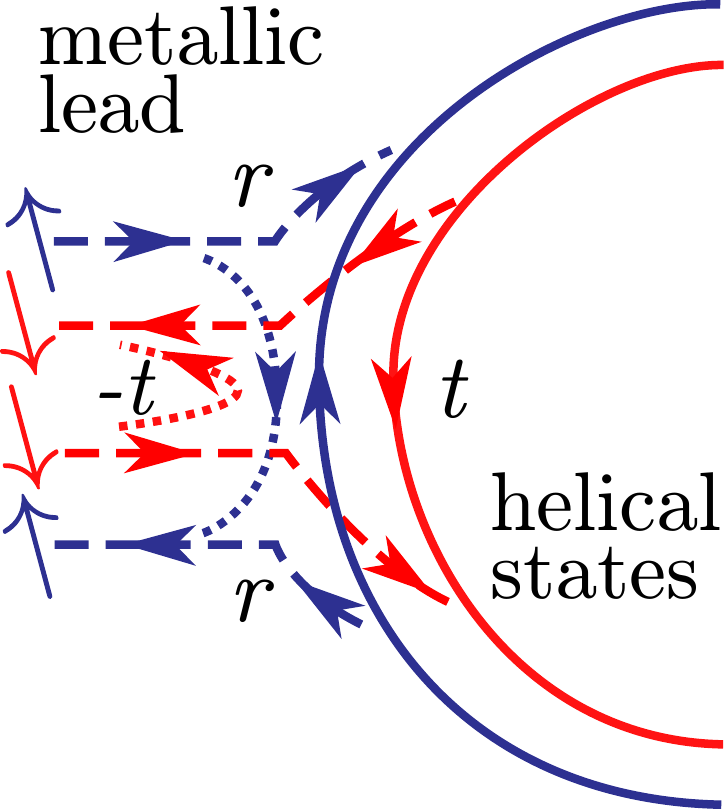}
\caption{\label{fig:pointcontact}
Point contact between the helical edge state
and the spinful wire. Blue (red) lines correspond to spin up (down) electrons. Solid lines depict trajectories inside the interferometer. Dashed lines 
show incoming and outgoing electron trajectories. Dotted lines illustrate reflection by contact.
}
\end{figure}

We suppose that HES are tunnel-coupled to metallic leads.  
These leads are modeled 
by single-channel spinful wires so that   electrons are injected into  the helical states through the so-called tunnel Y junctions. Different spin projections do not mix at the tunneling contacts so that   electrons entering the edge with opposite spins move in the opposite directions (see Fig.~\ref{fig:pointcontact}). Such contacts are characterized by two amplitudes $r$ and $t,$  obeying $|t|^2+|r|^2=1.$
We assume that $t$  and  $r$ are real   and   positive and
 parameterize them as follows:
\be
  r=\sqrt{1-e^{-2\lambda}}, \quad
 t=e^{-\lambda},
  \quad 0< \lambda<\infty.
\ee

\section{Calculation of conductance and polarization}
The transmission coefficient, $\mathcal T$, the spin transmission coefficient, $\mathcal T_s$, and the spin polarization, $\mathcal P = \mathcal T_s / \mathcal T $, are expressed via the fractions of transmitted electrons, $\mathcal T_\alpha$,  with spin projection  $\alpha =\uparrow,\downarrow$. 
Introducing  transfer matrix $\hat t$   of the interferormeter as a whole, we get
\begin{align}
\mathcal T &= 
\tfrac12  ( {\mathcal T_\uparrow + \mathcal T_\downarrow} ) = 
\tfrac12 \left \langle {\rm Tr} \left(\hat t  \hat t^\dagger \right) \right \rangle_\epsilon \,, 
\\
\mathcal T_s&=\tfrac12 \left( \mathcal T_{ \uparrow}-\mathcal T_{ \downarrow} \right)
=  \tfrac12 \left \langle {\rm Tr} \left( \hat t \sigma_z  \hat t^\dagger \right)\right \rangle_\epsilon \,,
\end{align}
where the thermal averaging, $\langle \cdots \rangle_\epsilon= -\int d\epsilon (\cdots) \p_\epsilon f_F(\epsilon),$ is performed with the Fermi function  $f_F(\epsilon).$   Here  we assume that the incoming electrons  are unpolarized.  The tunneling  conductance of this setup
 is given by
\begin{equation}
G=2\frac{e^2}{h}\mathcal{T} \, , 
\end{equation}
where factor $2$ accounts for two conducting channels.

The transfer matrix $\hat t$  is defined as follows
 \be 
 \begin{pmatrix} 
 a^\uparrow  \\
 a^\downarrow 
\end{pmatrix} = \hat t      \begin{pmatrix} 
b^\uparrow  \\
 b^ \downarrow \end{pmatrix},
 \ee
where  $(b^\uparrow, b^\downarrow)$  and   $(a^\uparrow, a^\downarrow)$ are the amplitudes of incoming (from the left contact) and outgoing (from the right contact) waves, respectively (see Fig.~\ref{fig:transfermatrix}).  Transfer  matrix corresponding to $  \hat S_M$ reads
\be \hat M= \frac{1}{\cos \theta}   \begin{pmatrix}                                                 1 &  i \sin \theta e^{i \xi} \\
                                                -i \sin \theta e^{-i \xi} & 1 
                                    \end{pmatrix}
                              ,\ee
where $\xi= \varphi- 2 k x_0$ and $k$ is the electron momentum.   The matrix $\hat t$ is expressed 
 in terms of $\hat   M$ as follows \cite{Niyazov2020}
\aleq{
\hat t=&  \frac{r^2 e^{2 \pi i \phi L_1/L}}{t} \begin{pmatrix} e^{i k L_1} & 0 \\ 0&e^{-i k L_1} \end{pmatrix} \begin{pmatrix} t & 0 \\ 0&1 \end{pmatrix}\hat g \begin{pmatrix} 1 & 0 \\ 0&1/t \end{pmatrix}, \\
 \hat g=& \frac{1}{1- e^{2 \pi i \phi}\hat M \begin{pmatrix} t^2 e^{i k L} & 0 \\ 0&e^{-i k L}/t^2 \end{pmatrix}} \hat M  \begin{pmatrix} 1 & 0 \\ 0&-1 \end{pmatrix}.
}
The matrix $\hat g$ can be  represented as follows
\be
\hat g  = \cos \theta \left[  \begin{pmatrix} 0 & 0 \\ 0 & -1   \end{pmatrix} + \sum\limits_{\alpha=\pm}   \frac{1+ \alpha \hat H}{1-t^2 e^{i( kL + \alpha  2\pi \phi_0)}}  \right],
\label{expansionH}
\ee
where $\phi_0$ obeys
\be
\cos(2\pi \phi_0)=\cos\theta \cos(2\pi \phi)\,,
\ee
and 
 \be \hat H =
\begin{pmatrix} a & b  e^{  i \xi}
\\ b e^{ - i \xi}  & -a \end{pmatrix}.
\label{Hadamard}
\ee 
The coefficients
\begin{align}
&   
a= i\frac{
e^{-2 \pi i \phi} -\cos(2\pi \phi_0) \cos\theta}{\cos \theta
\sin(2 \pi \phi_0)}, 
\\
&
b= \frac{ e^{-2 \pi i \phi} \tan\theta}{ \sin (2\pi \phi_0) }
\end{align}
 obey $a^2+b^2=1$ and depend on the strength of the impurity and the magnetic flux only, while
 the dependence on the   energy is encoded in the exponents
 $e^{\pm i \xi}$ entering off-diagonal terms of $\hat H.$

The possibility to express the transmission amplitude $\hat t$ in terms of resonance denominators~\eqref{expansionH} is of primary importance for further high temperature averaging. It allows us to do exact thermal averaging for arbitrary magnetic impurity strength, in the distinction with previous calculations \cite{Niyazov2018,Niyazov2020}, where perturbative expansion over impurity strength was used for calculation of $\cal T$ and $\cal P$. 
We first note that the energy dependence in the transfer matrix $\hat t$ appears not only in the resonance denominators but also  in terms $e^{i k L_1}$ and $e^{i 2 k x_0}.$ However, 
for  relevant combinations $ {\rm Tr} \left(\hat t  \hat t^\dagger \right)$  and $ {\rm Tr} \left(\hat t  \sigma_z   \hat t^\dagger \right),$
all energy dependent terms in numerators cancel. It reflects the universality of the HES based interferometers. AB oscillations do not depend on details of the setup: position of the impurity, $x_0$, length of the shoulders, $L_{1,2}$, and the Berry phase, $\delta$, as thoroughly discussed in Ref.~\cite{Niyazov2018}. Thus, we average only the following combinations 
\aleq{
&\left \langle \frac{1}{1-t^2 e^{i( kL + \alpha  2\pi \phi_0)}}\frac{1}{1-t^2 e^{-i( kL + \beta  2\pi \phi_0)}}\right \rangle_\epsilon \\
&\qquad =\frac{1}{1-t^4 e^{(\alpha-\beta)2 \pi \phi_0}},
\\ 
&\left \langle \frac{1}{1-t^2 e^{\pm i( kL + \alpha  2\pi \phi_0)}} \right \rangle_\epsilon=1.
}
Using these formulas, the straightforward algebraic calculation yields
\aleq{\label{eq:TPexact}
\mathcal T &=\tanh \lambda \left (1 -\frac{\sin ^2\theta   \sinh ^2\lambda  \cosh (2 \lambda) }{   \cosh ^2(2 \lambda) -\cos ^2\theta  \cos ^2(2 \pi  \phi) } \right ) , \\
\mathcal T_s & =-\frac{\sin ^2\theta   \sinh^2  \lambda   \cosh (2 \lambda)}{ 
 \cosh ^2(2 \lambda) -\cos ^2\theta  \cos ^2(2 \pi  \phi)
 } \, , \\ 
\mathcal P 
&= 
- 
\frac{ \tanh  \lambda \sin ^2\theta  }{1+  
 \cos ^2\theta  \left(\tanh ^2\lambda    - \frac{\cos ^2(2 \pi  \phi)}{\cosh ^2\lambda\cosh (2 \lambda )} \right)  
 } \,  .
}
 
This is the main result of the current paper.
We emphasize that these expressions are valid for arbitrary   
 tunneling amplitude of the contacts,  arbitrary  strength  of the magnetic impurity, and for any  
magnetic flux.  

\begin{figure}
\centerline{\includegraphics[width=\linewidth]{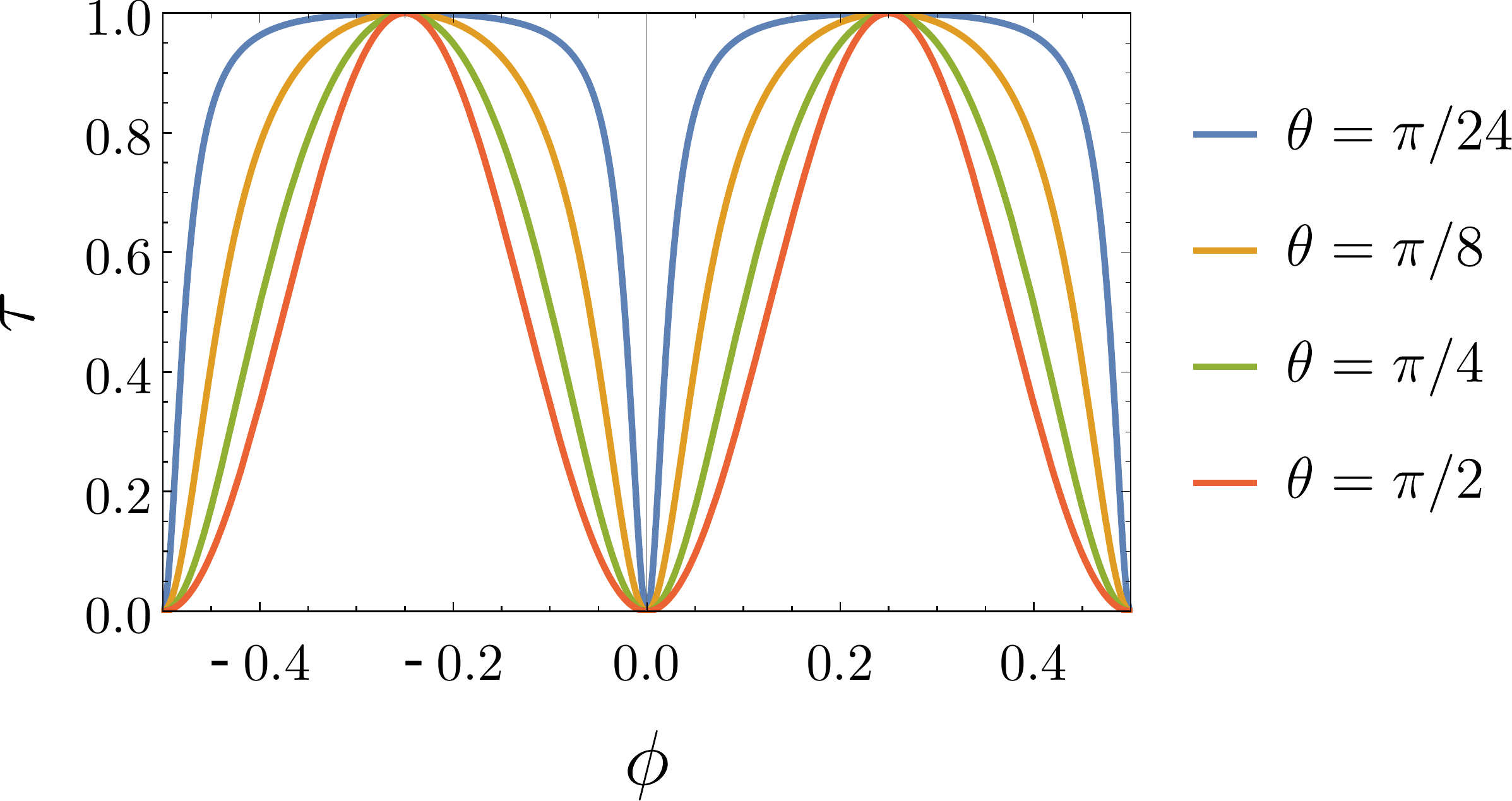}}
\caption{ Sharp antiresonances in the normalized conductance $\tau(\phi)$, Eq.\  \eqref{gPhi}, for different values of the    magnetic impurity strength.
  }
\label{Gnorm}
\end{figure}

We see that the transmission coefficient has minima at $\phi = n/2$, and maxima at $\phi = 1/4 + n/2$ with integer $n$.  Instead of  $\mathcal T (\phi)$ it is
convenient to  introduce the following normalized function
\aleq{
\tau (\phi)  &= \frac{ \mathcal T (\phi)  -  \mathcal T (0)} {\mathcal T (1/4)  - \mathcal T (0)} \\ &= \frac{\cosh ^2 (2 \lambda) \sin ^2(2 \pi  \phi) }{\cosh ^2 (2 \lambda)-\cos ^2 \theta  \cos ^2 (2 \pi  \phi) } \,.
\label{gPhi}
}
which is plotted in Fig.~\ref{Gnorm} for four different values of the magnetic impurity strength, $\theta$. Sharp antiresonances structure of $\mathcal T(\phi)$ is transforming to an oscillation shape with increasing of $\theta$. Simultaneously, depth of conductance antiresonances is decreasing such that $\mathcal T(\phi) \to \text{const}$ for  $\theta  \to \pi/2$. This case corresponds to the ideal reflection of electrons on the impurity. 

Fourier spectrum of conductance oscillations is  convenient representation for the analysis of experimental data~\cite{Ziegler2018,Savchenko2019}. Remarkably, that Fourier coefficients of the transmissions coefficient, $\mathcal{T}^{(n)}=\int d\phi\, \mathcal T(\phi) e^{i 4 \pi \phi n}, $ obey universal relation
\aleq{
\frac{\mathcal{T}^{(n)}}{\mathcal{T}^{(n+1)}}=-1 + 2 \kappa(\kappa-\sqrt{\kappa^2-1}) \, ,
}
where $\kappa=\cosh (2 \lambda)/ \cos \theta$.

An interesting relation between the  transmission coefficients $\mathcal T_{ \alpha} $ can be noticed both in the exact quantum result \eqref{eq:TPexact} and in its classical counterpart, \eqref{eq:Tclas}. While the values of $\mathcal T_{ \alpha}$   depend on the strength of the magnetic impurity and the flux, one observes that the property 
\[ \mathcal T_{ \uparrow} + e^{2\lambda} \mathcal T_{ \downarrow} =e^{2\lambda}-1 ,\]
involves only the transparency of the contact, $t=e^{-\lambda}$. It is tempting to regard this property as general one, but further inspection reveals that it holds only for the impurity in the ``upper'' shoulder of the ring in Fig.\ \ref{fig:transfermatrix}, while for the impurity in the lower part of the ring we should interchange $\mathcal T_{ \uparrow} \leftrightarrow \mathcal T_{ \downarrow}$ in the above formula. For impurities in both shoulders of AB ring the above relation is also violated, which can be checked rather easily for the classical trajectories, using the formulas from \cite{Niyazov2020}.

Let us now analyze the limiting cases. 

\subsection{Open interferometer}
 For the open interferometer, $\lambda \to \infty$, equations~\eqref{eq:TPexact} read
\aleq{
\mathcal{T}&= \frac{1+\cos^2 \theta}{2} \,, \\
\mathcal{P}&= - \frac{\sin^2 \theta}{2 - \sin^2 \theta} \,.
}
Two possible transmission channels have a trivial contribution to $\mathcal{T}$: spin-down channel conduct electrons without loss, whereas electrons scatter on the magnetic impurity in the spin-up channel with forwarding scattering amplitude $\cos \theta$  (see Fig.~\ref{fig:transfermatrix}). For full reflection case, we have $\theta=\pi /2$, $\mathcal{T}=1/2$ and outgoing electrons are fully polarized, $\mathcal{P}=-1$. This is a classical result which is insensitive to dephasing.

\subsection{Almost closed interferometer}
For the almost closed interferometer, $\lambda \to 0$, the interference contributions play important role:
\aleq{
\mathcal{T}&= \lambda - \frac{\lambda^3 \sin^2 \theta}{1+4\lambda^2 - \cos^2 \theta \cos^2 (2 \pi \phi)} \,, \\
\mathcal{P}&= - \frac{\lambda \sin^2 \theta}{1+\lambda^2(4-\sin^2 \theta) - \cos^2 \theta \cos^2 (2 \pi \phi)} \,.
}
We see that sharp antiresonances appear at half-integer and integer values of the flux, $\phi$ by contrast to conventional interferometers,  where only half-integer resonances exist~\cite{Dmitriev2015}.   The difference is related to the absence of backscattering by non-magnetic contacts in the case of helical edge states~\cite{Niyazov2018}.

\subsection{Weak magnetic impurity}
 For the previously studied case~\cite{Niyazov2018,Niyazov2020} of   weak scattering on the magnetic impurity, $\theta \to 0$, we obtain
\aleq{
\mathcal{T}&= \tanh \lambda \left(1 - \frac{2 \theta^2 \cosh (2 \lambda) \sinh^2 \lambda}{\cosh (4\lambda) -\cos (4 \pi \phi)} \right) \,, \\
\mathcal{P}&= - \frac{\theta^2}{2}\frac{ \sinh (4 \lambda)}{\cosh (4\lambda) -\cos (4 \pi \phi)} \,.
}

\section{Quantum flux-independent  processes}

\begin{figure}
\includegraphics[width=0.7\columnwidth]{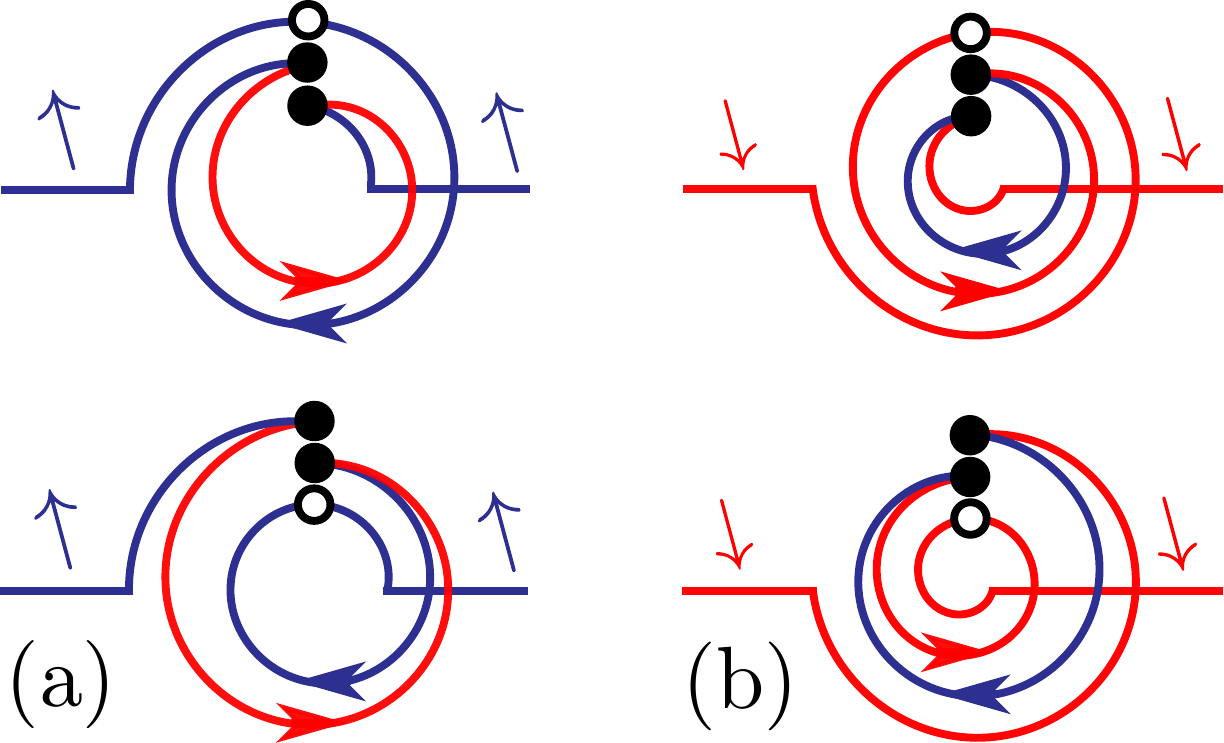}
\caption{
Simplest quantum processes which do not sensitive to magnetic flux.}
\label{nophase}
 \end{figure}

Let us now discuss one interesting aspect of our central result \eqref{eq:TPexact}, namely, the possible recovery of the classical contribution  upon the  averaging over the magnetic flux. 
Previously we have shown ~\cite{Niyazov2018,Niyazov2020} that the classical result was correctly reproduced by such averaging when keeping the terms of order $\theta^2$. 
The exact expressions for the  classical result was found there as 
  \aleq{
  \label{eq:Tclas}
 \mathcal{T}_\text{cl}&= \tanh \lambda \left(1 - \frac12\frac{\tanh^2 \lambda  \tan^2\theta}{1+\tan^2 \theta \coth (2\lambda)}\right) \,, \\
 \mathcal T_{s,\text{cl}}&= - \frac12\frac{\tanh^2 \lambda \tan^2\theta}{1+\tan^2 \theta \coth (2\lambda)} \,.
 }
Now we  perform the  averaging over the magnetic flux of our  quantum result  \eqref{eq:TPexact} and  obtain 
 \aleq{
 \langle\mathcal{T}\rangle_\phi&= \tanh \lambda \left(1 - \frac{\sqrt{2}\sin^2\theta \sinh^2 \lambda}{\sqrt{\cosh (4\lambda) -\cos (2 \theta)}} \right) \,, \\
\langle \mathcal T_s 
\rangle_\phi&= - \frac{\sqrt{2}\sin^2\theta \sinh^2 \lambda}{\sqrt{\cosh (4\lambda) -\cos (2 \theta)}} \,.
 }
Clearly these expressions are different  and  subtracting the purely classical  result from the  quantum one, averaged over the magnetic flux, we get the non-zero result. It  implies  the existence of quantum flux-independent processes. They appear first in the order $\theta^4$:
\aleq{
 \mathcal{T}_\text{cl} -\langle\mathcal{T}\rangle_\phi=-\frac{t^8}{\left(1+t^2\right)^{4}}\theta ^4 + \mathcal{O}(\theta^6) \, .
}
The coefficient $t^8$ before $\theta^4$ means that 
the electron is passing  the contacts at least four times. 
The simplest  examples of such processes are shown in Fig.~\ref{nophase}: panel (a) for spin up $\sim t^8+\mathcal O (t^{10})$ and panel (b) for spin down $\sim t^{12}+\mathcal O (t^{12})$. 

\section{Conclusions}
We have studied high-temperature transport
through the helical  Aharonov-Bohm interferometer tunnel-coupled to metallic leads. We focused on effect induced by strong magnetic impurity placed in one arm of  the interferometer and demonstrated that  
the tunneling conductance and the spin polarization of the outgoing  electrons show sharp antiresonances at integer and half-integer values of the dimensionless flux $\phi$. 
We calculated the spin-dependent transmission coefficients, $\mathcal T_{ \alpha}$,  
for arbitrary values of the tunneling coupling and   the  magnetic impurity strength.  We generalize previously obtained  results, describing  transport near the  resonant values of $\phi$, to the arbitrary value of magnetic flux. We also discussed  special quantum  effects which do not show up for weak impurity.     

 We found optimal conditions for spin filtering. Specifically, we demonstrated that spin polarization of the outgoing electrons reach  100\%  in the limit of  strong magnetic impurity and  open interferometer. The conductance of the setup equals in this case
$e^2/h.$
In this limit all quantum effects are suppressed and  the transmission through the interferometer has purely classical nature, i.e. is robust to dephasing. 

To conclude, a helical AB interferometer with strong magnetic impurity   allows to create large spin polarization. Remarkably, such polarization can be reached  even for $\phi=0,$ i.e. without a magnetic field. The 
scattering strength can be controlled by in-plane magnetic field or by external ferromagnetic tip, providing additional ways to manipulate spin polarization. 
These features add up to the remarkable properties of topological materials, making them even more attractive for spintronics, magnetic field detection, quantum networking, and quantum computing.

\section{Acknowledgements}

The calculation of spin polarization (R.N.) was funded by RFBR, project number 19-32-60077. The calculation of conductance (D.A. and V.K.) was funded by 
 the Russian Science Foundation, Grant No. 20-12-00147.  The work of R.N. and V.K. was partially supported by  Foundation for the Advancement  of Theoretical Physics and Mathematics.


\begin{thebibliography}{38}%
\makeatletter
\providecommand \@ifxundefined [1]{%
 \@ifx{#1\undefined}
}%
\providecommand \@ifnum [1]{%
 \ifnum #1\expandafter \@firstoftwo
 \else \expandafter \@secondoftwo
 \fi
}%
\providecommand \@ifx [1]{%
 \ifx #1\expandafter \@firstoftwo
 \else \expandafter \@secondoftwo
 \fi
}%
\providecommand \natexlab [1]{#1}%
\providecommand \enquote  [1]{``#1''}%
\providecommand \bibnamefont  [1]{#1}%
\providecommand \bibfnamefont [1]{#1}%
\providecommand \citenamefont [1]{#1}%
\providecommand \href@noop [0]{\@secondoftwo}%
\providecommand \href [0]{\begingroup \@sanitize@url \@href}%
\providecommand \@href[1]{\@@startlink{#1}\@@href}%
\providecommand \@@href[1]{\endgroup#1\@@endlink}%
\providecommand \@sanitize@url [0]{\catcode `\\12\catcode `\$12\catcode
  `\&12\catcode `\#12\catcode `\^12\catcode `\_12\catcode `\%12\relax}%
\providecommand \@@startlink[1]{}%
\providecommand \@@endlink[0]{}%
\providecommand \url  [0]{\begingroup\@sanitize@url \@url }%
\providecommand \@url [1]{\endgroup\@href {#1}{\urlprefix }}%
\providecommand \urlprefix  [0]{URL }%
\providecommand \Eprint [0]{\href }%
\providecommand \doibase [0]{http://dx.doi.org/}%
\providecommand \selectlanguage [0]{\@gobble}%
\providecommand \bibinfo  [0]{\@secondoftwo}%
\providecommand \bibfield  [0]{\@secondoftwo}%
\providecommand \translation [1]{[#1]}%
\providecommand \BibitemOpen [0]{}%
\providecommand \bibitemStop [0]{}%
\providecommand \bibitemNoStop [0]{.\EOS\space}%
\providecommand \EOS [0]{\spacefactor3000\relax}%
\providecommand \BibitemShut  [1]{\csname bibitem#1\endcsname}%
\let\auto@bib@innerbib\@empty
\bibitem [{\citenamefont {Bernevig}\ and\ \citenamefont
  {Hughes}(2013)}]{Bernevig2013}%
  \BibitemOpen
  \bibfield  {author} {\bibinfo {author} {\bibfnamefont {B.}~\bibnamefont
  {Bernevig}}\ and\ \bibinfo {author} {\bibfnamefont {T.}~\bibnamefont
  {Hughes}},\ }\href {https://books.google.ru/books?id=wOn7JHSSxrsC} {\emph
  {\bibinfo {title} {Topological Insulators and Topological Superconductors}}}\
  (\bibinfo  {publisher} {Princeton University Press},\ \bibinfo {year}
  {2013})\BibitemShut {NoStop}%
\bibitem [{\citenamefont {Hasan}\ and\ \citenamefont {Kane}(2010)}]{Hasan2010}%
  \BibitemOpen
  \bibfield  {author} {\bibinfo {author} {\bibfnamefont {M.~Z.}\ \bibnamefont
  {Hasan}}\ and\ \bibinfo {author} {\bibfnamefont {C.~L.}\ \bibnamefont
  {Kane}},\ }\href {\doibase 10.1103/RevModPhys.82.3045} {\bibfield  {journal}
  {\bibinfo  {journal} {Rev. Mod. Phys.}\ }\textbf {\bibinfo {volume} {82}},\
  \bibinfo {pages} {3045} (\bibinfo {year} {2010})}\BibitemShut {NoStop}%
\bibitem [{\citenamefont {Qi}\ and\ \citenamefont {Zhang}(2011)}]{Qi2011}%
  \BibitemOpen
  \bibfield  {author} {\bibinfo {author} {\bibfnamefont {X.-L.}\ \bibnamefont
  {Qi}}\ and\ \bibinfo {author} {\bibfnamefont {S.-C.}\ \bibnamefont {Zhang}},\
  }\href {\doibase 10.1103/RevModPhys.83.1057} {\bibfield  {journal} {\bibinfo
  {journal} {Rev. Mod. Phys.}\ }\textbf {\bibinfo {volume} {83}},\ \bibinfo
  {pages} {1057} (\bibinfo {year} {2011})}\BibitemShut {NoStop}%
\bibitem [{\citenamefont {Kane}\ and\ \citenamefont {Mele}(2005)}]{Kane2005}%
  \BibitemOpen
  \bibfield  {author} {\bibinfo {author} {\bibfnamefont {C.~L.}\ \bibnamefont
  {Kane}}\ and\ \bibinfo {author} {\bibfnamefont {E.~J.}\ \bibnamefont
  {Mele}},\ }\href {\doibase 10.1103/PhysRevLett.95.226801} {\bibfield
  {journal} {\bibinfo  {journal} {Phys. Rev. Lett.}\ }\textbf {\bibinfo
  {volume} {95}},\ \bibinfo {pages} {226801} (\bibinfo {year}
  {2005})}\BibitemShut {NoStop}%
\bibitem [{\citenamefont {Bernevig}\ \emph {et~al.}(2006)\citenamefont
  {Bernevig}, \citenamefont {Hughes},\ and\ \citenamefont {Zhang}}]{predicted}%
  \BibitemOpen
  \bibfield  {author} {\bibinfo {author} {\bibfnamefont {B.~A.}\ \bibnamefont
  {Bernevig}}, \bibinfo {author} {\bibfnamefont {T.~L.}\ \bibnamefont
  {Hughes}}, \ and\ \bibinfo {author} {\bibfnamefont {S.-C.}\ \bibnamefont
  {Zhang}},\ }\href {\doibase 10.1126/science.1133734} {\bibfield  {journal}
  {\bibinfo  {journal} {Science}\ }\textbf {\bibinfo {volume} {314}},\ \bibinfo
  {pages} {1757} (\bibinfo {year} {2006})}\BibitemShut {NoStop}%
\bibitem [{\citenamefont {K\"onig}\ \emph {et~al.}(2007)\citenamefont
  {K\"onig}, \citenamefont {Wiedmann}, \citenamefont {Brune}, \citenamefont
  {Roth}, \citenamefont {Buhmann}, \citenamefont {Molenkamp}, \citenamefont
  {Qi},\ and\ \citenamefont {Zhang}}]{confirmed}%
  \BibitemOpen
  \bibfield  {author} {\bibinfo {author} {\bibfnamefont {M.}~\bibnamefont
  {K\"onig}}, \bibinfo {author} {\bibfnamefont {S.}~\bibnamefont {Wiedmann}},
  \bibinfo {author} {\bibfnamefont {C.}~\bibnamefont {Brune}}, \bibinfo
  {author} {\bibfnamefont {A.}~\bibnamefont {Roth}}, \bibinfo {author}
  {\bibfnamefont {H.}~\bibnamefont {Buhmann}}, \bibinfo {author} {\bibfnamefont
  {L.~W.}\ \bibnamefont {Molenkamp}}, \bibinfo {author} {\bibfnamefont {X.-L.}\
  \bibnamefont {Qi}}, \ and\ \bibinfo {author} {\bibfnamefont {S.-C.}\
  \bibnamefont {Zhang}},\ }\href {\doibase 10.1126/science.1148047} {\bibfield
  {journal} {\bibinfo  {journal} {Science}\ }\textbf {\bibinfo {volume}
  {318}},\ \bibinfo {pages} {766} (\bibinfo {year} {2007})}\BibitemShut
  {NoStop}%
\bibitem [{\citenamefont {Roth}\ \emph {et~al.}(2009)\citenamefont {Roth},
  \citenamefont {Br{\"u}ne}, \citenamefont {Buhmann}, \citenamefont
  {Molenkamp}, \citenamefont {Maciejko}, \citenamefont {Qi},\ and\
  \citenamefont {Zhang}}]{Roth2009}%
  \BibitemOpen
  \bibfield  {author} {\bibinfo {author} {\bibfnamefont {A.}~\bibnamefont
  {Roth}}, \bibinfo {author} {\bibfnamefont {C.}~\bibnamefont {Br{\"u}ne}},
  \bibinfo {author} {\bibfnamefont {H.}~\bibnamefont {Buhmann}}, \bibinfo
  {author} {\bibfnamefont {L.~W.}\ \bibnamefont {Molenkamp}}, \bibinfo {author}
  {\bibfnamefont {J.}~\bibnamefont {Maciejko}}, \bibinfo {author}
  {\bibfnamefont {X.-L.}\ \bibnamefont {Qi}}, \ and\ \bibinfo {author}
  {\bibfnamefont {S.-C.}\ \bibnamefont {Zhang}},\ }\href {\doibase
  10.1126/science.1174736} {\bibfield  {journal} {\bibinfo  {journal}
  {Science}\ }\textbf {\bibinfo {volume} {325}},\ \bibinfo {pages} {294}
  (\bibinfo {year} {2009})}\BibitemShut {NoStop}%
\bibitem [{\citenamefont {Gusev}\ \emph {et~al.}(2011)\citenamefont {Gusev},
  \citenamefont {Kvon}, \citenamefont {Shegai}, \citenamefont {Mikhailov},
  \citenamefont {Dvoretsky},\ and\ \citenamefont {Portal}}]{Gusev2011}%
  \BibitemOpen
  \bibfield  {author} {\bibinfo {author} {\bibfnamefont {G.~M.}\ \bibnamefont
  {Gusev}}, \bibinfo {author} {\bibfnamefont {Z.~D.}\ \bibnamefont {Kvon}},
  \bibinfo {author} {\bibfnamefont {O.~A.}\ \bibnamefont {Shegai}}, \bibinfo
  {author} {\bibfnamefont {N.~N.}\ \bibnamefont {Mikhailov}}, \bibinfo {author}
  {\bibfnamefont {S.~A.}\ \bibnamefont {Dvoretsky}}, \ and\ \bibinfo {author}
  {\bibfnamefont {J.~C.}\ \bibnamefont {Portal}},\ }\href {\doibase
  10.1103/PhysRevB.84.121302} {\bibfield  {journal} {\bibinfo  {journal} {Phys.
  Rev. B}\ }\textbf {\bibinfo {volume} {84}},\ \bibinfo {pages} {121302}
  (\bibinfo {year} {2011})}\BibitemShut {NoStop}%
\bibitem [{\citenamefont {Brüne}\ \emph {et~al.}(2012)\citenamefont {Brüne},
  \citenamefont {Roth}, \citenamefont {Buhmann}, \citenamefont {Hankiewicz},
  \citenamefont {Molenkamp}, \citenamefont {Maciejko}, \citenamefont {Qi},\
  and\ \citenamefont {Zhang}}]{Brune2012}%
  \BibitemOpen
  \bibfield  {author} {\bibinfo {author} {\bibfnamefont {C.}~\bibnamefont
  {Brüne}}, \bibinfo {author} {\bibfnamefont {A.}~\bibnamefont {Roth}},
  \bibinfo {author} {\bibfnamefont {H.}~\bibnamefont {Buhmann}}, \bibinfo
  {author} {\bibfnamefont {E.~M.}\ \bibnamefont {Hankiewicz}}, \bibinfo
  {author} {\bibfnamefont {L.~W.}\ \bibnamefont {Molenkamp}}, \bibinfo {author}
  {\bibfnamefont {J.}~\bibnamefont {Maciejko}}, \bibinfo {author}
  {\bibfnamefont {X.-L.}\ \bibnamefont {Qi}}, \ and\ \bibinfo {author}
  {\bibfnamefont {S.-C.}\ \bibnamefont {Zhang}},\ }\href@noop {} {\bibfield
  {journal} {\bibinfo  {journal} {Nature Physics}\ }\textbf {\bibinfo {volume}
  {8}},\ \bibinfo {pages} {485} (\bibinfo {year} {2012})}\BibitemShut {NoStop}%
\bibitem [{\citenamefont {Kononov}\ \emph {et~al.}(2015)\citenamefont
  {Kononov}, \citenamefont {Egorov}, \citenamefont {Kvon}, \citenamefont
  {Mikhailov}, \citenamefont {Dvoretsky},\ and\ \citenamefont
  {Deviatov}}]{Kononov2015}%
  \BibitemOpen
  \bibfield  {author} {\bibinfo {author} {\bibfnamefont {A.}~\bibnamefont
  {Kononov}}, \bibinfo {author} {\bibfnamefont {S.~V.}\ \bibnamefont {Egorov}},
  \bibinfo {author} {\bibfnamefont {Z.~D.}\ \bibnamefont {Kvon}}, \bibinfo
  {author} {\bibfnamefont {N.~N.}\ \bibnamefont {Mikhailov}}, \bibinfo {author}
  {\bibfnamefont {S.~A.}\ \bibnamefont {Dvoretsky}}, \ and\ \bibinfo {author}
  {\bibfnamefont {E.~V.}\ \bibnamefont {Deviatov}},\ }\href {\doibase
  10.1134/S0021364015120115} {\bibfield  {journal} {\bibinfo  {journal} {JETP
  Letters}\ }\textbf {\bibinfo {volume} {101}},\ \bibinfo {pages} {814}
  (\bibinfo {year} {2015})}\BibitemShut {NoStop}%
\bibitem [{\citenamefont {Peng}\ \emph {et~al.}(2010)\citenamefont {Peng},
  \citenamefont {Lai}, \citenamefont {Kong}, \citenamefont {Meister},
  \citenamefont {Chen}, \citenamefont {Qi}, \citenamefont {Zhang},
  \citenamefont {Shen},\ and\ \citenamefont {Cui}}]{Peng2010}%
  \BibitemOpen
  \bibfield  {author} {\bibinfo {author} {\bibfnamefont {H.}~\bibnamefont
  {Peng}}, \bibinfo {author} {\bibfnamefont {K.}~\bibnamefont {Lai}}, \bibinfo
  {author} {\bibfnamefont {D.}~\bibnamefont {Kong}}, \bibinfo {author}
  {\bibfnamefont {S.}~\bibnamefont {Meister}}, \bibinfo {author} {\bibfnamefont
  {Y.}~\bibnamefont {Chen}}, \bibinfo {author} {\bibfnamefont {X.-L.}\
  \bibnamefont {Qi}}, \bibinfo {author} {\bibfnamefont {S.-C.}\ \bibnamefont
  {Zhang}}, \bibinfo {author} {\bibfnamefont {Z.-X.}\ \bibnamefont {Shen}}, \
  and\ \bibinfo {author} {\bibfnamefont {Y.}~\bibnamefont {Cui}},\ }\href
  {\doibase 10.1038/nmat2609} {\bibfield  {journal} {\bibinfo  {journal} {Nat
  Mater}\ }\textbf {\bibinfo {volume} {9}},\ \bibinfo {pages} {225} (\bibinfo
  {year} {2010})}\BibitemShut {NoStop}%
\bibitem [{\citenamefont {Lin}\ \emph {et~al.}(2017)\citenamefont {Lin},
  \citenamefont {Wang}, \citenamefont {Wang}, \citenamefont {Li}, \citenamefont
  {Li}, \citenamefont {Yu},\ and\ \citenamefont {Liao}}]{Lin2017}%
  \BibitemOpen
  \bibfield  {author} {\bibinfo {author} {\bibfnamefont {B.-C.}\ \bibnamefont
  {Lin}}, \bibinfo {author} {\bibfnamefont {S.}~\bibnamefont {Wang}}, \bibinfo
  {author} {\bibfnamefont {L.-X.}\ \bibnamefont {Wang}}, \bibinfo {author}
  {\bibfnamefont {C.-Z.}\ \bibnamefont {Li}}, \bibinfo {author} {\bibfnamefont
  {J.-G.}\ \bibnamefont {Li}}, \bibinfo {author} {\bibfnamefont
  {D.}~\bibnamefont {Yu}}, \ and\ \bibinfo {author} {\bibfnamefont {Z.-M.}\
  \bibnamefont {Liao}},\ }\href {\doibase 10.1103/PhysRevB.95.235436}
  {\bibfield  {journal} {\bibinfo  {journal} {Phys. Rev. B}\ }\textbf {\bibinfo
  {volume} {95}},\ \bibinfo {pages} {235436} (\bibinfo {year}
  {2017})}\BibitemShut {NoStop}%
\bibitem [{\citenamefont {Bardarson}\ \emph {et~al.}(2010)\citenamefont
  {Bardarson}, \citenamefont {Brouwer},\ and\ \citenamefont
  {Moore}}]{Bardarson2010}%
  \BibitemOpen
  \bibfield  {author} {\bibinfo {author} {\bibfnamefont {J.~H.}\ \bibnamefont
  {Bardarson}}, \bibinfo {author} {\bibfnamefont {P.~W.}\ \bibnamefont
  {Brouwer}}, \ and\ \bibinfo {author} {\bibfnamefont {J.~E.}\ \bibnamefont
  {Moore}},\ }\href {\doibase 10.1103/PhysRevLett.105.156803} {\bibfield
  {journal} {\bibinfo  {journal} {Phys. Rev. Lett.}\ }\textbf {\bibinfo
  {volume} {105}},\ \bibinfo {pages} {156803} (\bibinfo {year}
  {2010})}\BibitemShut {NoStop}%
\bibitem [{\citenamefont {Bardarson}\ and\ \citenamefont
  {Moore}(2013)}]{Bardarson2013}%
  \BibitemOpen
  \bibfield  {author} {\bibinfo {author} {\bibfnamefont {J.~H.}\ \bibnamefont
  {Bardarson}}\ and\ \bibinfo {author} {\bibfnamefont {J.~E.}\ \bibnamefont
  {Moore}},\ }\href {\doibase 10.1088/0034-4885/76/5/056501} {\bibfield
  {journal} {\bibinfo  {journal} {Reports on Progress in Physics}\ }\textbf
  {\bibinfo {volume} {76}},\ \bibinfo {pages} {056501} (\bibinfo {year}
  {2013})}\BibitemShut {NoStop}%
\bibitem [{\citenamefont {Delplace}\ \emph {et~al.}(2012)\citenamefont
  {Delplace}, \citenamefont {Li},\ and\ \citenamefont
  {B\"uttiker}}]{Buttiker2012}%
  \BibitemOpen
  \bibfield  {author} {\bibinfo {author} {\bibfnamefont {P.}~\bibnamefont
  {Delplace}}, \bibinfo {author} {\bibfnamefont {J.}~\bibnamefont {Li}}, \ and\
  \bibinfo {author} {\bibfnamefont {M.}~\bibnamefont {B\"uttiker}},\ }\href
  {\doibase 10.1103/PhysRevLett.109.246803} {\bibfield  {journal} {\bibinfo
  {journal} {Phys. Rev. Lett.}\ }\textbf {\bibinfo {volume} {109}},\ \bibinfo
  {pages} {246803} (\bibinfo {year} {2012})}\BibitemShut {NoStop}%
\bibitem [{\citenamefont {Gusev}\ \emph {et~al.}(2015)\citenamefont {Gusev},
  \citenamefont {Kvon}, \citenamefont {Shegai}, \citenamefont {Mikhailov},\
  and\ \citenamefont {Dvoretsky}}]{KvonAB2015}%
  \BibitemOpen
  \bibfield  {author} {\bibinfo {author} {\bibfnamefont {G.}~\bibnamefont
  {Gusev}}, \bibinfo {author} {\bibfnamefont {Z.}~\bibnamefont {Kvon}},
  \bibinfo {author} {\bibfnamefont {O.}~\bibnamefont {Shegai}}, \bibinfo
  {author} {\bibfnamefont {N.}~\bibnamefont {Mikhailov}}, \ and\ \bibinfo
  {author} {\bibfnamefont {S.}~\bibnamefont {Dvoretsky}},\ }\href {\doibase
  https://doi.org/10.1016/j.ssc.2014.12.017} {\bibfield  {journal} {\bibinfo
  {journal} {Solid State Communications}\ }\textbf {\bibinfo {volume} {205}},\
  \bibinfo {pages} {4 } (\bibinfo {year} {2015})}\BibitemShut {NoStop}%
\bibitem [{\citenamefont {Chu}\ \emph {et~al.}(2009)\citenamefont {Chu},
  \citenamefont {Li}, \citenamefont {Jain},\ and\ \citenamefont
  {Shen}}]{Chu2009}%
  \BibitemOpen
  \bibfield  {author} {\bibinfo {author} {\bibfnamefont {R.-L.}\ \bibnamefont
  {Chu}}, \bibinfo {author} {\bibfnamefont {J.}~\bibnamefont {Li}}, \bibinfo
  {author} {\bibfnamefont {J.~K.}\ \bibnamefont {Jain}}, \ and\ \bibinfo
  {author} {\bibfnamefont {S.-Q.}\ \bibnamefont {Shen}},\ }\href {\doibase
  10.1103/PhysRevB.80.081102} {\bibfield  {journal} {\bibinfo  {journal} {Phys.
  Rev. B}\ }\textbf {\bibinfo {volume} {80}},\ \bibinfo {pages} {081102}
  (\bibinfo {year} {2009})}\BibitemShut {NoStop}%
\bibitem [{\citenamefont {Masuda}\ and\ \citenamefont
  {Kuramoto}(2012)}]{Masuda2012}%
  \BibitemOpen
  \bibfield  {author} {\bibinfo {author} {\bibfnamefont {S.}~\bibnamefont
  {Masuda}}\ and\ \bibinfo {author} {\bibfnamefont {Y.}~\bibnamefont
  {Kuramoto}},\ }\href {\doibase 10.1103/PhysRevB.85.195327} {\bibfield
  {journal} {\bibinfo  {journal} {Phys. Rev. B}\ }\textbf {\bibinfo {volume}
  {85}},\ \bibinfo {pages} {195327} (\bibinfo {year} {2012})}\BibitemShut
  {NoStop}%
\bibitem [{\citenamefont {Dutta}\ \emph {et~al.}(2016)\citenamefont {Dutta},
  \citenamefont {Saha},\ and\ \citenamefont {Jayannavar}}]{Dutta2016}%
  \BibitemOpen
  \bibfield  {author} {\bibinfo {author} {\bibfnamefont {P.}~\bibnamefont
  {Dutta}}, \bibinfo {author} {\bibfnamefont {A.}~\bibnamefont {Saha}}, \ and\
  \bibinfo {author} {\bibfnamefont {A.~M.}\ \bibnamefont {Jayannavar}},\ }\href
  {\doibase 10.1103/physrevb.94.195414} {\bibfield  {journal} {\bibinfo
  {journal} {Physical Review B}\ }\textbf {\bibinfo {volume} {94}},\ \bibinfo
  {pages} {195414} (\bibinfo {year} {2016})}\BibitemShut {NoStop}%
\bibitem [{\citenamefont {Konig}\ \emph {et~al.}(2007)\citenamefont {Konig},
  \citenamefont {Wiedmann}, \citenamefont {Brune}, \citenamefont {Roth},
  \citenamefont {Buhmann}, \citenamefont {Molenkamp}, \citenamefont {Qi},\ and\
  \citenamefont {Zhang}}]{Konig2007}%
  \BibitemOpen
  \bibfield  {author} {\bibinfo {author} {\bibfnamefont {M.}~\bibnamefont
  {Konig}}, \bibinfo {author} {\bibfnamefont {S.}~\bibnamefont {Wiedmann}},
  \bibinfo {author} {\bibfnamefont {C.}~\bibnamefont {Brune}}, \bibinfo
  {author} {\bibfnamefont {A.}~\bibnamefont {Roth}}, \bibinfo {author}
  {\bibfnamefont {H.}~\bibnamefont {Buhmann}}, \bibinfo {author} {\bibfnamefont
  {L.~W.}\ \bibnamefont {Molenkamp}}, \bibinfo {author} {\bibfnamefont {X.-L.}\
  \bibnamefont {Qi}}, \ and\ \bibinfo {author} {\bibfnamefont {S.-C.}\
  \bibnamefont {Zhang}},\ }\href {\doibase 10.1126/science.1148047} {\bibfield
  {journal} {\bibinfo  {journal} {Science}\ }\textbf {\bibinfo {volume}
  {318}},\ \bibinfo {pages} {766} (\bibinfo {year} {2007})}\BibitemShut
  {NoStop}%
\bibitem [{\citenamefont {Knez}\ \emph {et~al.}(2011)\citenamefont {Knez},
  \citenamefont {Du},\ and\ \citenamefont {Sullivan}}]{Knez2011}%
  \BibitemOpen
  \bibfield  {author} {\bibinfo {author} {\bibfnamefont {I.}~\bibnamefont
  {Knez}}, \bibinfo {author} {\bibfnamefont {R.-R.}\ \bibnamefont {Du}}, \ and\
  \bibinfo {author} {\bibfnamefont {G.}~\bibnamefont {Sullivan}},\ }\href
  {\doibase 10.1103/PhysRevLett.107.136603} {\bibfield  {journal} {\bibinfo
  {journal} {Phys. Rev. Lett.}\ }\textbf {\bibinfo {volume} {107}},\ \bibinfo
  {pages} {136603} (\bibinfo {year} {2011})}\BibitemShut {NoStop}%
\bibitem [{\citenamefont {Wu}\ \emph {et~al.}(2018)\citenamefont {Wu},
  \citenamefont {Fatemi}, \citenamefont {Gibson}, \citenamefont {Watanabe},
  \citenamefont {Taniguchi}, \citenamefont {Cava},\ and\ \citenamefont
  {Jarillo-Herrero}}]{Wu2018}%
  \BibitemOpen
  \bibfield  {author} {\bibinfo {author} {\bibfnamefont {S.}~\bibnamefont
  {Wu}}, \bibinfo {author} {\bibfnamefont {V.}~\bibnamefont {Fatemi}}, \bibinfo
  {author} {\bibfnamefont {Q.~D.}\ \bibnamefont {Gibson}}, \bibinfo {author}
  {\bibfnamefont {K.}~\bibnamefont {Watanabe}}, \bibinfo {author}
  {\bibfnamefont {T.}~\bibnamefont {Taniguchi}}, \bibinfo {author}
  {\bibfnamefont {R.~J.}\ \bibnamefont {Cava}}, \ and\ \bibinfo {author}
  {\bibfnamefont {P.}~\bibnamefont {Jarillo-Herrero}},\ }\href {\doibase
  10.1126/science.aan6003} {\bibfield  {journal} {\bibinfo  {journal}
  {Science}\ }\textbf {\bibinfo {volume} {359}},\ \bibinfo {pages} {76}
  (\bibinfo {year} {2018})}\BibitemShut {NoStop}%
\bibitem [{\citenamefont {Reis}\ \emph {et~al.}(2017)\citenamefont {Reis},
  \citenamefont {Li}, \citenamefont {Dudy}, \citenamefont {Bauernfeind},
  \citenamefont {Glass}, \citenamefont {Hanke}, \citenamefont {Thomale},
  \citenamefont {Sch{\"{a}}fer},\ and\ \citenamefont {Claessen}}]{Reis2017}%
  \BibitemOpen
  \bibfield  {author} {\bibinfo {author} {\bibfnamefont {F.}~\bibnamefont
  {Reis}}, \bibinfo {author} {\bibfnamefont {G.}~\bibnamefont {Li}}, \bibinfo
  {author} {\bibfnamefont {L.}~\bibnamefont {Dudy}}, \bibinfo {author}
  {\bibfnamefont {M.}~\bibnamefont {Bauernfeind}}, \bibinfo {author}
  {\bibfnamefont {S.}~\bibnamefont {Glass}}, \bibinfo {author} {\bibfnamefont
  {W.}~\bibnamefont {Hanke}}, \bibinfo {author} {\bibfnamefont
  {R.}~\bibnamefont {Thomale}}, \bibinfo {author} {\bibfnamefont
  {J.}~\bibnamefont {Sch{\"{a}}fer}}, \ and\ \bibinfo {author} {\bibfnamefont
  {R.}~\bibnamefont {Claessen}},\ }\href {\doibase 10.1126/science.aai8142}
  {\bibfield  {journal} {\bibinfo  {journal} {Science}\ }\textbf {\bibinfo
  {volume} {357}},\ \bibinfo {pages} {287} (\bibinfo {year}
  {2017})}\BibitemShut {NoStop}%
\bibitem [{\citenamefont {Li}\ \emph {et~al.}(2018)\citenamefont {Li},
  \citenamefont {Hanke}, \citenamefont {Hankiewicz}, \citenamefont {Reis},
  \citenamefont {Sch{\"{a}}fer}, \citenamefont {Claessen}, \citenamefont {Wu},\
  and\ \citenamefont {Thomale}}]{Li2018}%
  \BibitemOpen
  \bibfield  {author} {\bibinfo {author} {\bibfnamefont {G.}~\bibnamefont
  {Li}}, \bibinfo {author} {\bibfnamefont {W.}~\bibnamefont {Hanke}}, \bibinfo
  {author} {\bibfnamefont {E.~M.}\ \bibnamefont {Hankiewicz}}, \bibinfo
  {author} {\bibfnamefont {F.}~\bibnamefont {Reis}}, \bibinfo {author}
  {\bibfnamefont {J.}~\bibnamefont {Sch{\"{a}}fer}}, \bibinfo {author}
  {\bibfnamefont {R.}~\bibnamefont {Claessen}}, \bibinfo {author}
  {\bibfnamefont {C.}~\bibnamefont {Wu}}, \ and\ \bibinfo {author}
  {\bibfnamefont {R.}~\bibnamefont {Thomale}},\ }\href {\doibase
  10.1103/PhysRevB.98.165146} {\bibfield  {journal} {\bibinfo  {journal} {Phys.
  Rev. B}\ }\textbf {\bibinfo {volume} {98}},\ \bibinfo {pages} {165146}
  (\bibinfo {year} {2018})}\BibitemShut {NoStop}%
\bibitem [{\citenamefont {St{\"{u}}hler}\ \emph {et~al.}(2020)\citenamefont
  {St{\"{u}}hler}, \citenamefont {Reis}, \citenamefont {M{\"{u}}ller},
  \citenamefont {Helbig}, \citenamefont {Schwemmer}, \citenamefont {Thomale},
  \citenamefont {Sch{\"{a}}fer},\ and\ \citenamefont {Claessen}}]{Stuhler2019}%
  \BibitemOpen
  \bibfield  {author} {\bibinfo {author} {\bibfnamefont {R.}~\bibnamefont
  {St{\"{u}}hler}}, \bibinfo {author} {\bibfnamefont {F.}~\bibnamefont {Reis}},
  \bibinfo {author} {\bibfnamefont {T.}~\bibnamefont {M{\"{u}}ller}}, \bibinfo
  {author} {\bibfnamefont {T.}~\bibnamefont {Helbig}}, \bibinfo {author}
  {\bibfnamefont {T.}~\bibnamefont {Schwemmer}}, \bibinfo {author}
  {\bibfnamefont {R.}~\bibnamefont {Thomale}}, \bibinfo {author} {\bibfnamefont
  {J.}~\bibnamefont {Sch{\"{a}}fer}}, \ and\ \bibinfo {author} {\bibfnamefont
  {R.}~\bibnamefont {Claessen}},\ }\href {\doibase 10.1038/s41567-019-0697-z}
  {\bibfield  {journal} {\bibinfo  {journal} {Nat. Phys.}\ }\textbf {\bibinfo
  {volume} {16}},\ \bibinfo {pages} {47} (\bibinfo {year} {2020})}\BibitemShut
  {NoStop}%
\bibitem [{\citenamefont {Jagla}\ and\ \citenamefont {Balseiro}(1993)}]{jagla}%
  \BibitemOpen
  \bibfield  {author} {\bibinfo {author} {\bibfnamefont {E.~A.}\ \bibnamefont
  {Jagla}}\ and\ \bibinfo {author} {\bibfnamefont {C.~A.}\ \bibnamefont
  {Balseiro}},\ }\href {\doibase 10.1103/PhysRevLett.70.639} {\bibfield
  {journal} {\bibinfo  {journal} {Phys. Rev. Lett.}\ }\textbf {\bibinfo
  {volume} {70}},\ \bibinfo {pages} {639} (\bibinfo {year} {1993})}\BibitemShut
  {NoStop}%
\bibitem [{\citenamefont {Dmitriev}\ \emph {et~al.}(2010)\citenamefont
  {Dmitriev}, \citenamefont {Gornyi}, \citenamefont {Kachorovskii},\ and\
  \citenamefont {Polyakov}}]{dmitriev}%
  \BibitemOpen
  \bibfield  {author} {\bibinfo {author} {\bibfnamefont {A.~P.}\ \bibnamefont
  {Dmitriev}}, \bibinfo {author} {\bibfnamefont {I.~V.}\ \bibnamefont
  {Gornyi}}, \bibinfo {author} {\bibfnamefont {V.~Y.}\ \bibnamefont
  {Kachorovskii}}, \ and\ \bibinfo {author} {\bibfnamefont {D.~G.}\
  \bibnamefont {Polyakov}},\ }\href {\doibase 10.1103/PhysRevLett.105.036402}
  {\bibfield  {journal} {\bibinfo  {journal} {Phys. Rev. Lett.}\ }\textbf
  {\bibinfo {volume} {105}},\ \bibinfo {pages} {036402} (\bibinfo {year}
  {2010})}\BibitemShut {NoStop}%
\bibitem [{\citenamefont {Shmakov}\ \emph {et~al.}(2013)\citenamefont
  {Shmakov}, \citenamefont {Dmitriev},\ and\ \citenamefont
  {Kachorovskii}}]{Shmakov2013}%
  \BibitemOpen
  \bibfield  {author} {\bibinfo {author} {\bibfnamefont {P.~M.}\ \bibnamefont
  {Shmakov}}, \bibinfo {author} {\bibfnamefont {A.~P.}\ \bibnamefont
  {Dmitriev}}, \ and\ \bibinfo {author} {\bibfnamefont {V.~Y.}\ \bibnamefont
  {Kachorovskii}},\ }\href {\doibase 10.1103/PhysRevB.87.235417} {\bibfield
  {journal} {\bibinfo  {journal} {Phys. Rev. B}\ }\textbf {\bibinfo {volume}
  {87}},\ \bibinfo {pages} {235417} (\bibinfo {year} {2013})}\BibitemShut
  {NoStop}%
\bibitem [{\citenamefont {Dmitriev}\ \emph {et~al.}(2015)\citenamefont
  {Dmitriev}, \citenamefont {Gornyi}, \citenamefont {Kachorovskii},
  \citenamefont {Polyakov},\ and\ \citenamefont {Shmakov}}]{Dmitriev2015}%
  \BibitemOpen
  \bibfield  {author} {\bibinfo {author} {\bibfnamefont {A.~P.}\ \bibnamefont
  {Dmitriev}}, \bibinfo {author} {\bibfnamefont {I.~V.}\ \bibnamefont
  {Gornyi}}, \bibinfo {author} {\bibfnamefont {V.~Y.}\ \bibnamefont
  {Kachorovskii}}, \bibinfo {author} {\bibfnamefont {D.~G.}\ \bibnamefont
  {Polyakov}}, \ and\ \bibinfo {author} {\bibfnamefont {P.~M.}\ \bibnamefont
  {Shmakov}},\ }\href {\doibase 10.1134/S0021364014240059} {\bibfield
  {journal} {\bibinfo  {journal} {JETP Letters}\ }\textbf {\bibinfo {volume}
  {100}},\ \bibinfo {pages} {839} (\bibinfo {year} {2015})}\BibitemShut
  {NoStop}%
\bibitem [{\citenamefont {Dmitriev}\ \emph {et~al.}(2017)\citenamefont
  {Dmitriev}, \citenamefont {Gornyi}, \citenamefont {Kachorovskii},\ and\
  \citenamefont {Polyakov}}]{SCS2017}%
  \BibitemOpen
  \bibfield  {author} {\bibinfo {author} {\bibfnamefont {A.~P.}\ \bibnamefont
  {Dmitriev}}, \bibinfo {author} {\bibfnamefont {I.~V.}\ \bibnamefont
  {Gornyi}}, \bibinfo {author} {\bibfnamefont {V.~Y.}\ \bibnamefont
  {Kachorovskii}}, \ and\ \bibinfo {author} {\bibfnamefont {D.~G.}\
  \bibnamefont {Polyakov}},\ }\href {\doibase 10.1103/PhysRevB.96.115417}
  {\bibfield  {journal} {\bibinfo  {journal} {Phys. Rev. B}\ }\textbf {\bibinfo
  {volume} {96}},\ \bibinfo {pages} {115417} (\bibinfo {year}
  {2017})}\BibitemShut {NoStop}%
\bibitem [{\citenamefont {Niyazov}\ \emph {et~al.}(2018)\citenamefont
  {Niyazov}, \citenamefont {Aristov},\ and\ \citenamefont
  {Kachorovskii}}]{Niyazov2018}%
  \BibitemOpen
  \bibfield  {author} {\bibinfo {author} {\bibfnamefont {R.~A.}\ \bibnamefont
  {Niyazov}}, \bibinfo {author} {\bibfnamefont {D.~N.}\ \bibnamefont
  {Aristov}}, \ and\ \bibinfo {author} {\bibfnamefont {V.~Y.}\ \bibnamefont
  {Kachorovskii}},\ }\href {\doibase 10.1103/PhysRevB.98.045418} {\bibfield
  {journal} {\bibinfo  {journal} {Phys. Rev. B}\ }\textbf {\bibinfo {volume}
  {98}},\ \bibinfo {pages} {045418} (\bibinfo {year} {2018})}\BibitemShut
  {NoStop}%
\bibitem [{\citenamefont {Niyazov}\ \emph {et~al.}(2020)\citenamefont
  {Niyazov}, \citenamefont {Aristov},\ and\ \citenamefont
  {Kachorovskii}}]{Niyazov2020}%
  \BibitemOpen
  \bibfield  {author} {\bibinfo {author} {\bibfnamefont {R.~A.}\ \bibnamefont
  {Niyazov}}, \bibinfo {author} {\bibfnamefont {D.~N.}\ \bibnamefont
  {Aristov}}, \ and\ \bibinfo {author} {\bibfnamefont {V.~Y.}\ \bibnamefont
  {Kachorovskii}},\ }\href {\doibase 10.1038/s41524-020-00442-z} {\bibfield
  {journal} {\bibinfo  {journal} {npj Comput. Mater.}\ }\textbf {\bibinfo
  {volume} {6}},\ \bibinfo {pages} {174} (\bibinfo {year} {2020})}\BibitemShut
  {NoStop}%
\bibitem [{\citenamefont {Du}\ \emph {et~al.}(2015)\citenamefont {Du},
  \citenamefont {Knez}, \citenamefont {Sullivan},\ and\ \citenamefont
  {Du}}]{Du2015}%
  \BibitemOpen
  \bibfield  {author} {\bibinfo {author} {\bibfnamefont {L.}~\bibnamefont
  {Du}}, \bibinfo {author} {\bibfnamefont {I.}~\bibnamefont {Knez}}, \bibinfo
  {author} {\bibfnamefont {G.}~\bibnamefont {Sullivan}}, \ and\ \bibinfo
  {author} {\bibfnamefont {R.-R.}\ \bibnamefont {Du}},\ }\href {\doibase
  10.1103/PhysRevLett.114.096802} {\bibfield  {journal} {\bibinfo  {journal}
  {Phys. Rev. Lett.}\ }\textbf {\bibinfo {volume} {114}},\ \bibinfo {pages}
  {096802} (\bibinfo {year} {2015})}\BibitemShut {NoStop}%
\bibitem [{\citenamefont {Zhang}\ \emph {et~al.}(2014)\citenamefont {Zhang},
  \citenamefont {Zhang},\ and\ \citenamefont {Shen}}]{Zhang2014}%
  \BibitemOpen
  \bibfield  {author} {\bibinfo {author} {\bibfnamefont {S.-B.}\ \bibnamefont
  {Zhang}}, \bibinfo {author} {\bibfnamefont {Y.-Y.}\ \bibnamefont {Zhang}}, \
  and\ \bibinfo {author} {\bibfnamefont {S.-Q.}\ \bibnamefont {Shen}},\ }\href
  {\doibase 10.1103/PhysRevB.90.115305} {\bibfield  {journal} {\bibinfo
  {journal} {Phys. Rev. B}\ }\textbf {\bibinfo {volume} {90}},\ \bibinfo
  {pages} {115305} (\bibinfo {year} {2014})}\BibitemShut {NoStop}%
\bibitem [{\citenamefont {Hu}\ \emph {et~al.}(2016)\citenamefont {Hu},
  \citenamefont {Xu}, \citenamefont {Zhang},\ and\ \citenamefont
  {Zhou}}]{Hu2016a}%
  \BibitemOpen
  \bibfield  {author} {\bibinfo {author} {\bibfnamefont {L.-H.}\ \bibnamefont
  {Hu}}, \bibinfo {author} {\bibfnamefont {D.-H.}\ \bibnamefont {Xu}}, \bibinfo
  {author} {\bibfnamefont {F.-C.}\ \bibnamefont {Zhang}}, \ and\ \bibinfo
  {author} {\bibfnamefont {Y.}~\bibnamefont {Zhou}},\ }\href {\doibase
  10.1103/PhysRevB.94.085306} {\bibfield  {journal} {\bibinfo  {journal} {Phys.
  Rev. B}\ }\textbf {\bibinfo {volume} {94}},\ \bibinfo {pages} {085306}
  (\bibinfo {year} {2016})}\BibitemShut {NoStop}%
\bibitem [{\citenamefont {Kurilovich}\ \emph {et~al.}(2017)\citenamefont
  {Kurilovich}, \citenamefont {Kurilovich}, \citenamefont {Burmistrov},\ and\
  \citenamefont {Goldstein}}]{kur3}%
  \BibitemOpen
  \bibfield  {author} {\bibinfo {author} {\bibfnamefont {P.~D.}\ \bibnamefont
  {Kurilovich}}, \bibinfo {author} {\bibfnamefont {V.~D.}\ \bibnamefont
  {Kurilovich}}, \bibinfo {author} {\bibfnamefont {I.~S.}\ \bibnamefont
  {Burmistrov}}, \ and\ \bibinfo {author} {\bibfnamefont {M.}~\bibnamefont
  {Goldstein}},\ }\href {\doibase 10.1134/S0021364017210020} {\bibfield
  {journal} {\bibinfo  {journal} {JETP Lett.}\ }\textbf {\bibinfo {volume}
  {106}},\ \bibinfo {pages} {593} (\bibinfo {year} {2017})}\BibitemShut
  {NoStop}%
\bibitem [{\citenamefont {Ziegler}\ \emph {et~al.}(2018)\citenamefont
  {Ziegler}, \citenamefont {Kozlovsky}, \citenamefont {Gorini}, \citenamefont
  {Liu}, \citenamefont {Weish{\"{a}}upl}, \citenamefont {Maier}, \citenamefont
  {Fischer}, \citenamefont {Kozlov}, \citenamefont {Kvon}, \citenamefont
  {Mikhailov}, \citenamefont {Dvoretsky}, \citenamefont {Richter},\ and\
  \citenamefont {Weiss}}]{Ziegler2018}%
  \BibitemOpen
  \bibfield  {author} {\bibinfo {author} {\bibfnamefont {J.}~\bibnamefont
  {Ziegler}}, \bibinfo {author} {\bibfnamefont {R.}~\bibnamefont {Kozlovsky}},
  \bibinfo {author} {\bibfnamefont {C.}~\bibnamefont {Gorini}}, \bibinfo
  {author} {\bibfnamefont {M.-H.}\ \bibnamefont {Liu}}, \bibinfo {author}
  {\bibfnamefont {S.}~\bibnamefont {Weish{\"{a}}upl}}, \bibinfo {author}
  {\bibfnamefont {H.}~\bibnamefont {Maier}}, \bibinfo {author} {\bibfnamefont
  {R.}~\bibnamefont {Fischer}}, \bibinfo {author} {\bibfnamefont {D.~A.}\
  \bibnamefont {Kozlov}}, \bibinfo {author} {\bibfnamefont {Z.~D.}\
  \bibnamefont {Kvon}}, \bibinfo {author} {\bibfnamefont {N.}~\bibnamefont
  {Mikhailov}}, \bibinfo {author} {\bibfnamefont {S.~A.}\ \bibnamefont
  {Dvoretsky}}, \bibinfo {author} {\bibfnamefont {K.}~\bibnamefont {Richter}},
  \ and\ \bibinfo {author} {\bibfnamefont {D.}~\bibnamefont {Weiss}},\ }\href
  {\doibase 10.1103/PhysRevB.97.035157} {\bibfield  {journal} {\bibinfo
  {journal} {Phys. Rev. B}\ }\textbf {\bibinfo {volume} {97}},\ \bibinfo
  {pages} {035157} (\bibinfo {year} {2018})}\BibitemShut {NoStop}%
\bibitem [{\citenamefont {Savchenko}\ \emph {et~al.}(2019)\citenamefont
  {Savchenko}, \citenamefont {Kozlov}, \citenamefont {Vasilev}, \citenamefont
  {Kvon}, \citenamefont {Mikhailov}, \citenamefont {Dvoretsky},\ and\
  \citenamefont {Kolesnikov}}]{Savchenko2019}%
  \BibitemOpen
  \bibfield  {author} {\bibinfo {author} {\bibfnamefont {M.~L.}\ \bibnamefont
  {Savchenko}}, \bibinfo {author} {\bibfnamefont {D.~A.}\ \bibnamefont
  {Kozlov}}, \bibinfo {author} {\bibfnamefont {N.~N.}\ \bibnamefont {Vasilev}},
  \bibinfo {author} {\bibfnamefont {Z.~D.}\ \bibnamefont {Kvon}}, \bibinfo
  {author} {\bibfnamefont {N.~N.}\ \bibnamefont {Mikhailov}}, \bibinfo {author}
  {\bibfnamefont {S.~A.}\ \bibnamefont {Dvoretsky}}, \ and\ \bibinfo {author}
  {\bibfnamefont {A.~V.}\ \bibnamefont {Kolesnikov}},\ }\href {\doibase
  10.1103/PhysRevB.99.195423} {\bibfield  {journal} {\bibinfo  {journal} {Phys.
  Rev. B}\ }\textbf {\bibinfo {volume} {99}},\ \bibinfo {pages} {195423}
  (\bibinfo {year} {2019})}\BibitemShut {NoStop}%
\end{thebibliography}
\end{document}